\documentclass[
 reprint,
 amsmath,amssymb,
 aps,
longbibliography
]{revtex4-1}

\usepackage{graphicx}% Include figure files
\usepackage{dcolumn}% Align table columns on decimal point
\usepackage{bm}% bold math
\usepackage{hyperref}% add hypertext capabilities
\begin{document}

%\preprint{APS/123-QED}

\title{Fast approximate simulation of seismic waves with deep learning}% Force line breaks with \\

\author{B. Moseley}
\email{bmoseley@robots.ox.ac.uk}
\author{A. Markham}

\affiliation{Centre for Doctoral Training in Autonomous Intelligent Machines and Systems
\\ University of Oxford}

\author{T. Nissen-Meyer}

\affiliation{Department of Earth Sciences
\\ University of Oxford}%


%\affiliation{
%\affiliation{
%\affiliation{%


\date{\today}% It is always \today, today,

\begin{abstract}
We simulate the response of acoustic seismic waves in horizontally layered media using a deep neural network. In contrast to traditional finite-difference modelling techniques our network is able to directly approximate the recorded seismic response at multiple receiver locations in a single inference step, without needing to iteratively model the seismic wavefield through time. This results in an order of magnitude reduction in simulation time from the order of 1~s for FD modelling to the order of 0.1~s using our approach. Such a speed improvement could lead to real-time seismic simulation applications and benefit seismic inversion algorithms based on forward modelling, such as full waveform inversion. Our proof of concept deep neural network is trained using 50,000 synthetic examples of seismic waves propagating through different 2D horizontally layered velocity models. We discuss how our approach could be extended to arbitrary velocity models. Our deep neural network design is inspired by the WaveNet architecture used for speech synthesis. We also investigate using deep neural networks for simulating the full seismic wavefield and for carrying out seismic inversion directly.
\end{abstract}

\maketitle

%\tableofcontents

\section{\label{sec:intro}Introduction}

Seismic simulations are invaluable in many areas of geophysics.  In earthquake monitoring, they are a key tool for quantifying the ground motion of potential earthquakes \cite{earthquake}. In oil and gas prospecting, they are used to understand the seismic response of hydrocarbon reservoirs \cite{4dseismic, reservoirmodel1}. In geophysical surveying, they show how the subsurface is illuminated by different survey designs \cite{surveydesign}. In global geophysics, seismic simulations are invaluable for obtaining snapshots of the Earth's interior dynamics \cite{global} and for deciphering source or path effects from individual seismograms \cite{globalsim}.

Seismic simulations are heavily used in seismic inversion, which aims to estimate the unknown elastic properties of a medium given its seismic response \cite{inversion}. In Full Waveform Inversion (FWI), a strategy quickly becoming widespread in the field of seismic imaging, forward simulations are used thousands of times to iteratively estimate a medium's elastic properties \cite{fwi}.

Numerous methods exist for seismic simulation. The most prominent are Finite Difference (FD) and spectral element methods \cite{fd, sem}. They are able to capture a full range of relevant physics, including the effects of solid-fluid interfaces, intrinsic attenuation and anisotropy.

For both methods, the underlying wave equation is discretised to solve for the propagation of the full seismic wavefield. For an acoustic heterogeneous medium the wave equation is given by

\begin{equation}
\label{eq:waveeq}
\rho \nabla \cdot \left( {1 \over \rho } \nabla p \right) - {1 \over v^{2} } {\partial^{2} p \over \partial t^{2} } = - \rho {\partial^{2} i_{V} \over \partial t^{2} }~,
\end{equation}

where $p$ is the acoustic pressure, $i_{V}$ is a point source of volume injection (the seismic source), and $v = \sqrt{\kappa / \rho}$ is the velocity of the medium, with $\rho$ the density of the medium and $\kappa$ the adiabatic compression modulus \cite{waveeq}.

A major bottleneck when using seismic simulations is their computational cost. For example, FD modelling can involve millions of grid points and at each time step the wavefield must be iteratively updated over the entire grid. It is usual for large simulations to be implemented on supercomputer clusters and real-time simulation is typically not possible \cite{fastfd}. Reducing simulation time would benefit many  applications \cite{fdcomplexity}.

The field of deep learning has recently shown promise in its ability to make approximate predictions of physical phenomena. These approaches are able to learn about highly non-linear physics and some offer much faster inference times than traditional simulation \cite{deepcfd, earthquakeconv}.

There also exist a wealth of deep learning techniques applicable for synthesising time series data. The recent WaveNet network was able to synthesis speech from text inputs using a causally connected deep neural network \cite{wavenet}.

In this paper we present a faster, approximate and novel approach for the simulation of seismic waves using deep learning.  Instead of using traditional, iterative numerical methods to model the full wavefield, we predict the full seismic response directly at multiple receiver locations in a single inference step by using a deep neural network.

We use a modified WaveNet architecture for the network and train the network to predict the pressure response from seismic waves travelling through 2D horizontally layered acoustic media. Whilst we study simple layered velocity models as a proof of concept here, we discuss how our approach could be extended to simulate more arbitrary Earth models.

We will also show preliminary results from using deep convolutional neural networks to model the full seismic wavefield and a complementary WaveNet network to carry out seismic inversion directly.

\begin{figure}[ht]
\begin{center}
\includegraphics[width=8.5cm]{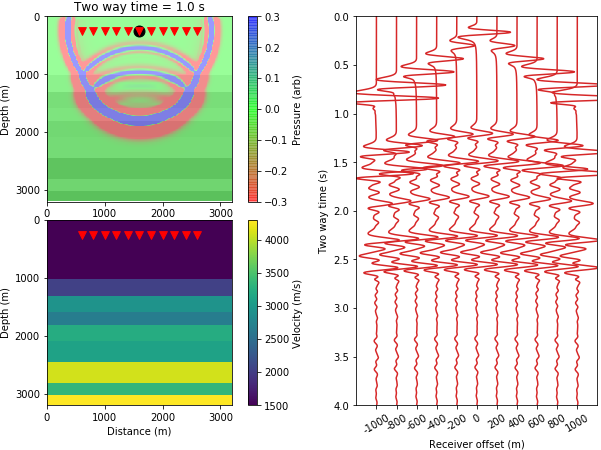}
\caption[]{Ground truth FD simulation example. Left, top: An 8~Hz Ricker seismic source is emitted close to the surface and propagates through a 2D horizontally layered acoustic Earth model. The black circle shows the source location. 11 receivers are placed at the same depth as the source with a horizontal spacing of 200~m (red triangles). The full wavefield pressure is overlain for a single snapshot in time ($t=1.0~\mathrm{s}$). Note seismic reflections occur at each velocity interface. Left, bottom: The Earth velocity model. The Earth model has a constant density of 2200~$\mathrm{kgm}^{-2}$. Right: The resulting ground truth pressure response recorded by each of the receivers, using FD modelling. A $t^2$ gain is applied to the receiver responses for display.}
\label{fig:example_simulation}
\end{center}
\end{figure}

\begin{figure}[ht]
\begin{center}
\includegraphics[width=8.5cm]{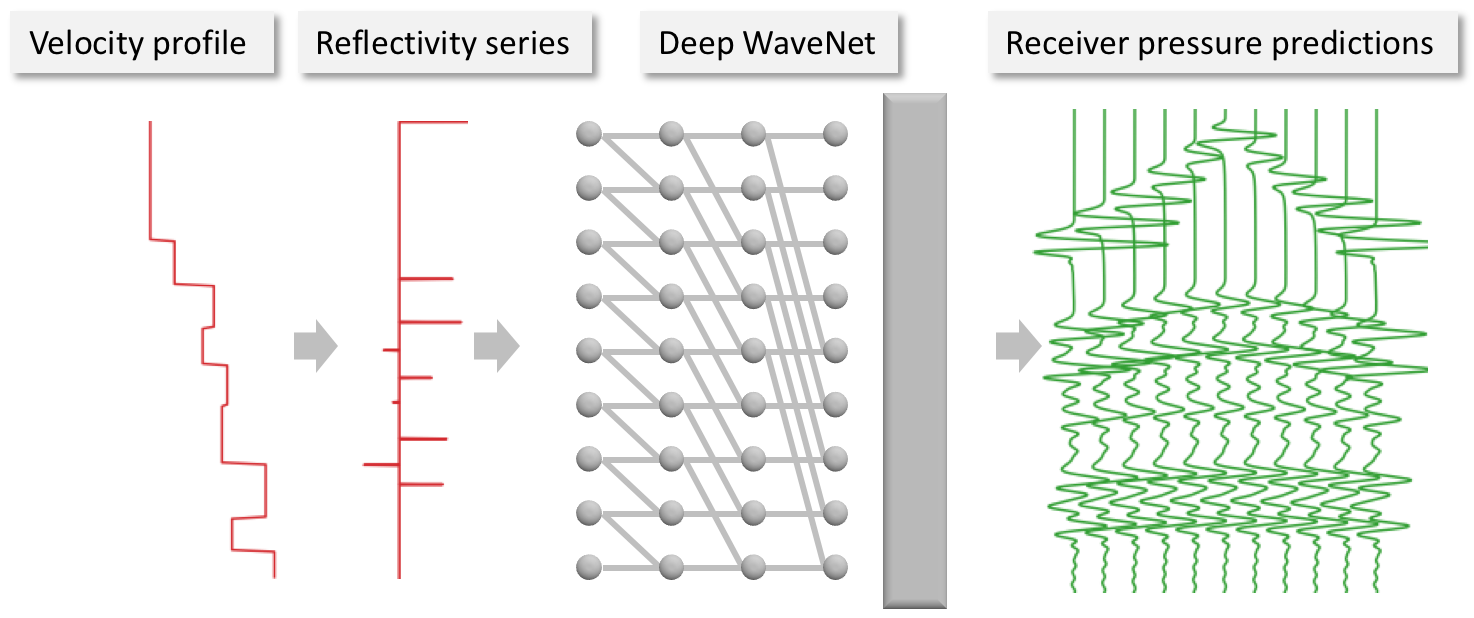}
\caption[]{Our WaveNet prediction workflow. Given a 1D Earth velocity profile as input (left), our WaveNet-inspired deep neural network (middle) outputs a prediction of the pressure responses at the 11 discrete receiver locations in Fig~\ref{fig:example_simulation} (right). The raw input 1D velocity profile sampled in depth is converted into its normal incidence reflectivity series sampled in time before being input into the network. The network is composed of 10 time-dilated causally-connected convolutional layers with a filter width of 2 and dilation rates which increase exponentially with layer depth. Each hidden layer of the network has same length as the input reflectivity series, 256 channels and a ReLU activation function. A final casually-connected convolutional layer with a filter width of 201 samples, 11 output channels and an identity activation is used to generate the output prediction.}
\label{fig:wavenet}
\end{center}
\end{figure}

\begin{figure}[ht]
\begin{center}
\includegraphics[width=7cm]{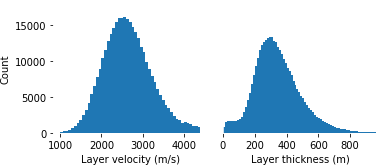}
\caption[]{Distribution of layer velocity and layer thickness over all examples in the training set.}
\label{fig:distributions}
\end{center}
\end{figure}

\begin{figure*}[t]
\begin{center}
\includegraphics[width=16.5cm]{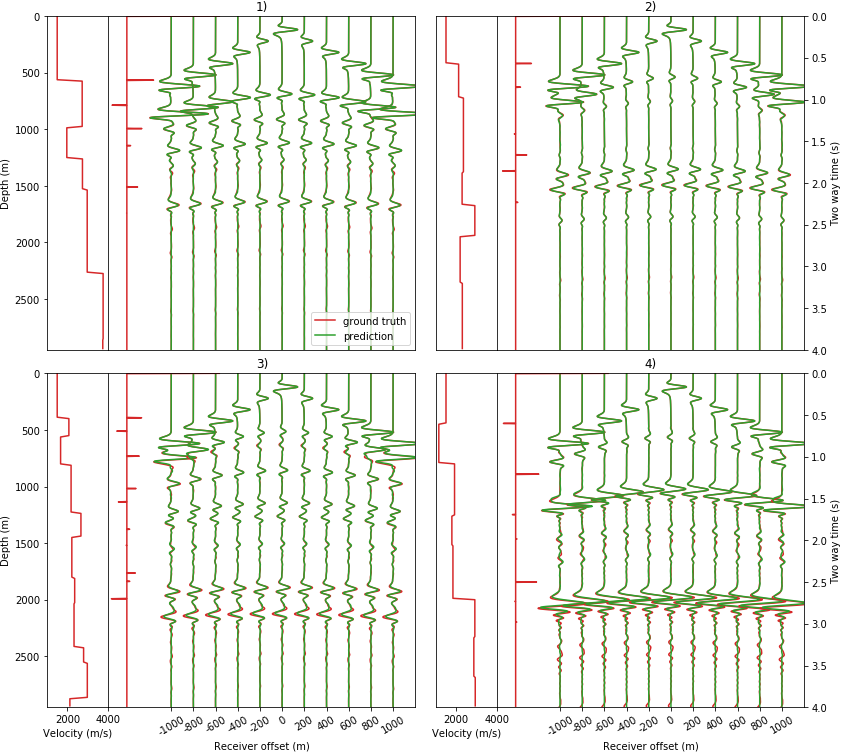}
\caption[]{WaveNet predictions for 4 randomly selected examples in our validation set. Red shows the input velocity model, its corresponding reflectivity series and the ground truth pressure response at the 11 receiver locations. Green shows the WaveNet prediction given the input reflectivity series for each example. A $t^2$ gain is applied to the receiver responses for display.}
\label{fig:11rec_result}
\end{center}
\end{figure*}

\begin{figure*}[t]
\begin{center}
\includegraphics[width=16.5cm]{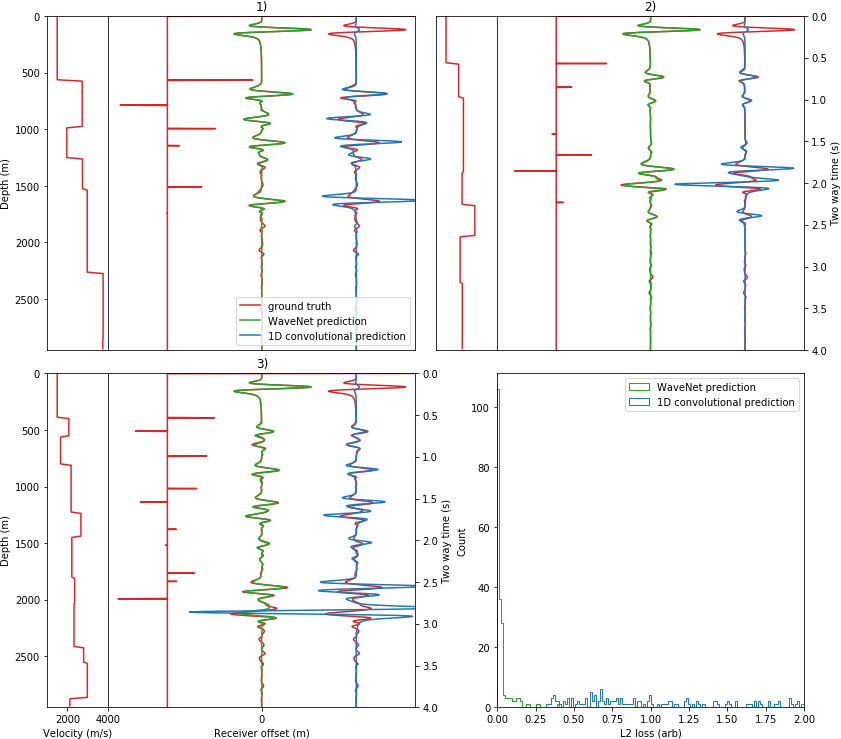}
\caption[]{Comparison of WaveNet prediction to 1D convolutional model. We compare our WaveNet prediction for 3 of the selected examples in Fig~\ref{fig:11rec_result} to a simple 1D convolutional model. Red shows the input velocity model, its corresponding reflectivity series and the ground truth pressure response at the zero-offset receiver. Green shows the WaveNet prediction at the zero-offset receiver and blue shows the 1D normal incidence convolutional model for the zero-offset receiver. Bottom right: the histogram of L2 loss values given by Eq.~\ref{eq:loss} for the zero-offset receiver prediction over our validation set of 200 examples. A $t^2$ gain is applied to the receiver responses for display.}
\label{fig:1dconv_result}
\end{center}
\end{figure*}

\subsection{\label{sec:related}Related Work}

Applying deep learning to physics problems is burgeoning field of research and there is much active work in this area. Lerer et al. \cite{deeptower} presented a deep convolutional network which could accurately predict whether randomly initialised wooden towers would fall or remain stable, given 2D images of the tower.

Guo et al. \cite{deepcfd} demonstrated that convolutional neural
networks could estimate flow fields in complex Computational Fluid Dynamics (CFD) calculations two orders of
magnitude faster than a traditional GPU-accelerated CFD solver. Their approach could allow real-time
feedback for aerodynamics applications.

Hooberman et al. \cite{deepparticle} presented deep learning methods for particle
classification, energy regression, and simulation for high-energy physics which could outperform traditional methods.

Geophysicists are also starting to use deep learning for seismic-related problems. Perol et al. \cite{earthquakeconv} presented an earthquake identification method using convolutional networks which is orders of magnitude faster than traditional techniques.

Weiqiang et al. \cite{stanfordwaves} presented a multi-scale convolutional network for predicting the evolution of the full seismic wavefield in heterogeneous density media. Their method was able to approximate the wavefield kinematics over multiple time steps, although it suffered from the accumulation of error over time. Krischer and Fichtner \cite{globaldeep} used a generative adversarial network to simulate seismograms from radially symmetric and smooth Earth models.

In seismic inversion, Araya-Polo et al. \cite{deeptomo} proposed a deep learning concept for carrying out seismic tomography using the semblance of common mid-point receiver gathers as input. Their method was able to make velocity model predictions from synthetic seismic data in a fraction of the time needed for traditional tomography techniques. In FWI, Richardson \cite{deepfwi1} demonstrated that a recurrent neural network framework with automatic differentiation can be used to carry out gradient calculations. Sun and  Demanet \cite{deepfwi2} showed a method for using deep learning to extrapolate low frequency seismic energy to improve the convergence of FWI algorithms.

\section{\label{sec:a}Fast seismic simulation using WaveNet}

\subsection{\label{sec:overview}Overview}

An example seismic simulation we wish to our deep neural network to learn is shown in Fig.~\ref{fig:example_simulation}. A point source is emitted in an Earth model and its pressure response is recorded by receivers placed at different locations in the model.

Our goal is to train a neural network which, given the Earth model as input, outputs an accurate prediction of the pressure response recorded at each of the receiver locations.

For this proof-of-principle study, we fix the receiver layout such that the receivers are horizontally displaced in 2D from the source. We only predict the acoustic pressure response from seismic waves travelling through 2D horizontally layered velocity models. We keep the density of the Earth model constant and use a fixed size Earth model. We expect the network to generalise well over unseen velocity models.

We will train our network using many ground truth examples of FD simulations. 

\subsection{\label{sec:wavenet}WaveNet architecture}

Our prediction workflow is summarised in Fig.~\ref{fig:wavenet}. The workflow consists of a preprocessing step, where we convert each input velocity model to its corresponding reflectivity series sampled in time (Fig.~\ref{fig:wavenet}, left), followed by a prediction step, where we use a deep neural network to predict the pressure response recorded by each receiver (Fig.~\ref{fig:wavenet}, middle). 

For horizontally layered velocity models and receivers horizontally offset from the source, each receiver pressure recording and the normal incidence reflectivity series of the input velocity model are causally correlated. Intuitively, a seismic reflection recorded after a short time has only travelled through a shallow part of the velocity model and its pressure response is at most dependent on past samples in the reflectivity series. 

Our prediction workflow honours this causal correlation by preprocessing the input velocity model into its corresponding reflectivity series and using a WaveNet-inspired network architecture with causally-connected convolutions to predict the receiver response.

We define the input velocity model to be a 1D profile of a horizontally layered Earth velocity model, with a depth of 3.2~km and a sample rate of 12.5~m. We convert the velocity profile to its corresponding normal incidence reflectivity series by carrying out a standard 1D depth-to-time conversion and inserting reflectivity values at normal incidence at each velocity interface, given by
\begin{equation}
\label{eq:r}
R = {\rho_{2}v_{2}-\rho_{1}v_{1} \over \rho_{2}v_{2}+\rho_{1}v_{1}}~,
\end{equation}
where $\rho_{1}$,$v_{1}$ and $\rho_{2}$,$v_{2}$ are the densities and velocities across the interface. The output reflectivity series has a length of 5~s and a sample rate of 4~ms. An example reflectivity series is shown in Fig.~\ref{fig:wavenet} (left).

Our WaveNet prediction network contains 10 causally-connected convolutional layers (Fig.~\ref{fig:wavenet}, middle). Each convolutional layer has the same length as the input reflectivity series, 256 hidden channels and a ReLU activation function. Similar to the original WaveNet work we use exponentially increasing dilation rates at each layer, which ensures that the first sample in the input reflectivity series is causally connected to the last sample of the output prediction.  We add a final casually-connected convolutional layer with 11 output channels and an identity activation to generate the output prediction, where each output channel corresponds to a receiver prediction.

\subsection{\label{sec:data}Training data generation}

To train the network, we use 50,000 synthetic ground truth example simulations generated by the open-source SEISMIC\_CPML library, which performs $2^{\mathrm{nd}}$-order acoustic FD modelling \cite{seiscpml}.

Each example simulation uses a horizontal layered 2D velocity model with an equal width and depth of 3.2~km and a sample rate of 12.5~m in both directions. (Fig.~\ref{fig:example_simulation}, bottom left). For all simulations we use a constant density model of 2200~$\mathrm{kgm}^{-2}$.

For each simulation the layer velocities and layer thickness are randomly sampled from Earth-realistic log-normal distributions. We add a small gradient randomly sampled from a normal distribution to each velocity model such that the velocity values tend to increase with depth, to be more Earth-realistic. The final distributions over layer velocities and layer thicknesses for the entire training set are shown in Fig.~\ref{fig:distributions}.

We use an 8~Hz Ricker source emitted close to the surface and record the pressure response at 11 receiver locations placed symmetrically around the source, horizontally offset every 200~m (Fig.~\ref{fig:example_simulation}, top left). The SEISMIC\_CPML library uses a convolutional perfectly matched layer boundary condition such that waves which reach the edge of the model are absorbed with negligible reflection.

We run each simulation for 5~s. We use a 2~ms sample rate to maintain accurate FD fidelity and downsample the resulting receiver pressure responses to 4~ms before using them in our prediction workflow.

We extract a  training example from each simulation, where a training example consists of the 1D layered velocity profile and the recorded pressure response at each of the 11 receivers. This gives a total of 50,000 training examples.

We also generate an independent test set of 10,000 examples to measure the generalisation performance of our network during training, using the same workflow.

\subsection{\label{sec:training}Training process}

We train the network using the Adam stochastic gradient descent algorithm  \cite{adam}. We use a standard L2 loss function with gain, given by

\begin{equation}
\label{eq:loss}
L = {1\over N} \lVert G(\hat{Y} - Y)\rVert_{2}^{2}~,
\end{equation}

where $\hat{Y}$ is the predicted receiver pressure response,  $Y$ is the ground truth receiver pressure response from FD modelling and $N$ is the number of training examples in each batch. The gain function $G$ has the fixed form $G=t^{2}$ where $t$ is the sample time.  This is added to empirically account for the spherical spreading of the wavefield  by increasing the weight of later time samples.

To help regularise the network during training we employ a dropout layer between the hidden output of our WaveNet architecture and the final convolutional layer, with a dropout rate of 0.4.

We use a learning rate of $1\mathrm{x}10^{-5}$, a batch size of 20 training examples and run training over 250,000 steps.

\subsection{\label{sec:results}Results}

During training both the test loss and the training loss converge to similar values, suggesting the network is able to generalise over different input velocity models.

To assess the performance of our trained network, we create a final validation set of 200 unseen examples. The receiver predictions for 4 randomly selected examples from this set are shown in Fig.~\ref{fig:11rec_result}.

A simple approximation to the receiver response at normal incidence is the 1D convolutional model, given by $\tilde{Y}=R * S$, where $R$ is the reflectivity series in time and $S$ is the source signature. We compare this 1D convolutional model with our WaveNet network predictions for the receiver at zero-offset  for 3 of the randomly selected examples in Fig.~\ref{fig:1dconv_result}.

For most time samples our network is able to accurately predict the amplitude of the recorded pressure response. Unlike the 1D convolutional model, it is also able to accurately predict the Normal Moveout (NMO) of the primary layer reflections with receiver offset, the amplitude and timing of the direct arrivals at the start of each receiver recording and the spherical spreading loss of the wavefield over time. 

We plot the histogram of the L2 loss values for the 1D convolutional model against our network prediction over the validation set in Fig.~\ref{fig:1dconv_result} (bottom right) and observe that our network has a significantly lower average loss of $0.03~\pm~0.05$ compared to $1.7~\pm~1.7$ for the 1D convolutional model.

Our network is able to convert the sparse reflectivity values into a frequency-limited seismic pressure response and in doing so implicitly learns the source signature.
 
The network struggles to model the multiple reverberations at the end of the receiver recordings in Fig.~\ref{fig:11rec_result}, perhaps because they have much more complex kinematics than the primary reflections.

We measure the average time taken to generate 11 receiver pressure responses using the SEISMIC\_CPML library over 100 runs on a single core of a 2.2 GHz Intel Core i7 processor to be $3.5 \pm 0.1$~s. Our network is able to generate a prediction of the 11 receiver responses in an average time of $0.18~\pm~0.01$~s using the same core. We note that the our prediction is easily parallelised using the Tensorflow \cite{tensorflow} framework; CPU multithreading on 8 cores allows us to reduce the prediction time to $0.054~\pm~0.007$~s and a Nvidia Tesla K80 GPU produces predictions with an average time of $0.015~\pm~0.001$~s.

\begin{figure}[ht]
\begin{center}
\includegraphics[width=8.5cm]{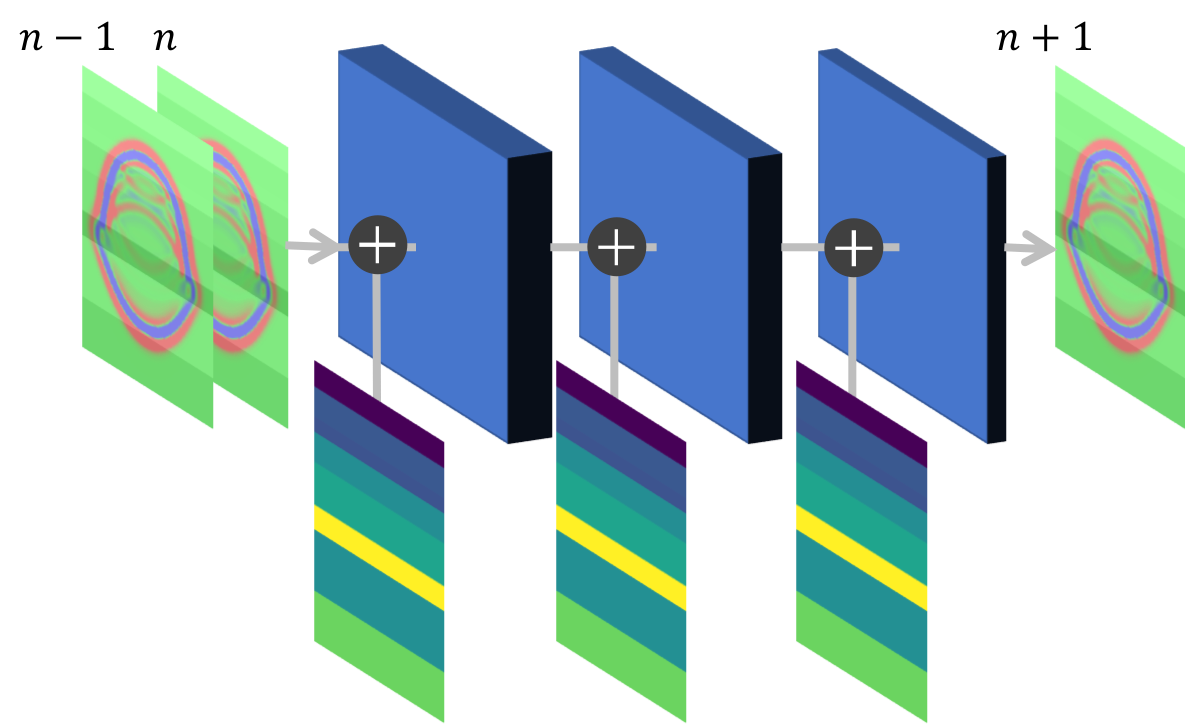}
\caption[]{Deep convolutional model used for full wavefield simulation. Our deep convolutional network predicts the $n+1$ wavefield frame given the two previous wavefield frames as input. The network has two hidden convolutional layers with filter size 5x5, output channel sizes of 128 and 32 and ReLU activation and a final output convolutional layer with filter size 5x5, 1 channel and an identity activation. The network is conditioned on the input velocity model by concatenating the velocity model to the input of each convolutional layer.}
\label{fig:convnet}
\end{center}
\end{figure}

\begin{figure*}[t]
\begin{center}
\includegraphics[width=16.5cm]{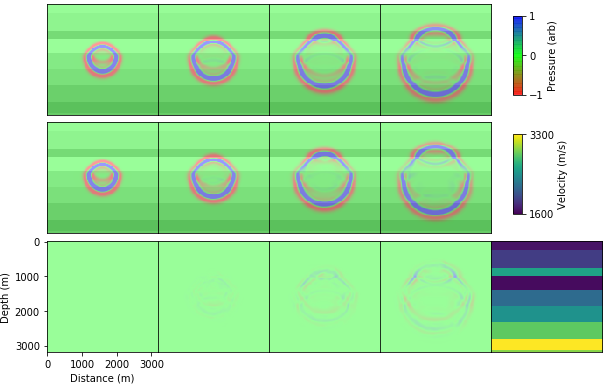}
\caption[]{Full wavefield simulation over time using our deep convolutional network. We recursively predict the evolution of an initial wavefield (left-most frame) in our validation set using our deep convolutional network (middle), compared to the ground truth FD modelling (top). We show the prediction at $t=0.00, 0.08, 0.16$ and $0.24$~s (left to right) and its difference to ground truth (bottom). The input velocity model is also shown (bottom right). }
\label{fig:wavefields_result}
\end{center}
\end{figure*}

\begin{figure*}[t]
\begin{center}
\includegraphics[width=16.5cm]{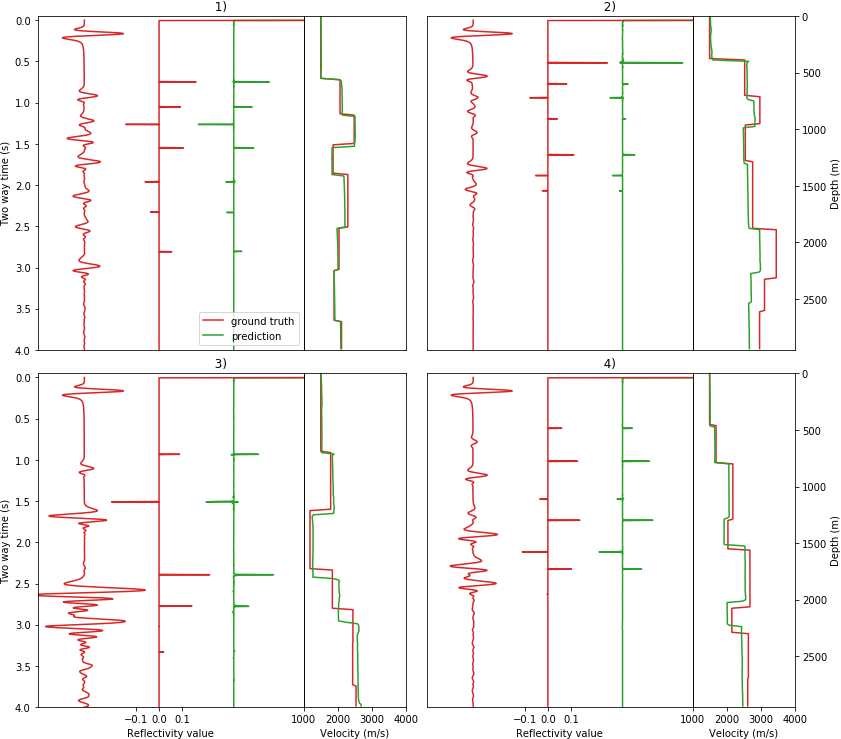}
\caption[]{Top: Inverse WaveNet predictions for 4 randomly selected examples in our seismic inversion validation set.  Red shows the input pressure response at the zero-offset receiver location, the ground truth reflectivity series and its corresponding velocity model.  Green shows the inverse WaveNet reflectivity series prediction and the resulting velocity prediction.}
\label{fig:inverse_result}
\end{center}
\end{figure*}

\section{\label{sec:b}Full wavefield simulation using deep convolutional networks}

\subsection{\label{sec:none}Overview}

In addition to simulating the pressure response at individual receiver locations, we carry out preliminary tests for using deep neural networks to simulate the full seismic wavefield.

Instead of predicting individual receiver responses we design a deep convolutional network which, given the two previous time steps of the pressure wavefield, predicts the next time step of the  wavefield over all points in space. This allows the network to be  used iteratively to predict the evolution of the full wavefield over multiple time steps, in a fashion similar to FD modelling.

Similar to Section~\ref{sec:a}, we only predict the 2D acoustic pressure response (Eq.~\ref{eq:waveeq}) and keep the density and the size of the Earth model fixed. We train the network to predict the wavefield evolution over time for different 2D horizontally layered velocity models and different starting wavefields as input.

\subsection{\label{sec:none}Deep convolutional model}

For an acoustic wave with constant density, the $2^{\mathrm{nd}}$-order finite difference update equation for the full wavefield (in 1D for brevity) is given by 
\begin{equation}
\label{eq:update}
p^{n+1}_{i} = -p^{n-1}_{i} + 2 p^{n}_{i} + C^2 ( p^{n}_{i+1} - 2 p^{n}_{i} + p^{n}_{i-1}  )   ~,
\end{equation}
where $p^{n}_{i}$ is the pressure at spatial sample $i$ and time sample $n$ and $C = v {\Delta t \over \Delta x}$ where $\Delta t$ is the time sample rate and $\Delta x$ is the spatial sample rate \cite{updateeq}. 

In both 1D and 2D the FD update equation only requires as input the current wavefield and the wavefield at the previous time step. Furthermore the updated wavefield at each sample location is only dependent on the current wavefield at neighbouring samples. 

Given these observations we use a deep convolutional neural network to approximate the update equation, shown in Fig.~\ref{fig:convnet}. The input to the network is the current and previous wavefield frames concatenated together and the output is a prediction of the wavefield at the next time step. The two input wavefields form a 236x236x2 input tensor. 

The convolutional network consists of 2 hidden 2D convolutional layers both with filters size of 5x5,  ReLU activations and output channel sizes of 128 and 32 respectively. A final convolutional layer with filter size 5x5, identity activation and 1 output channel is added for the output prediction.

We condition the network on the input 2D velocity model by concatenating the velocity model  to the input of each convolutional layer.

\subsection{\label{sec:none}Training process}

Training data is generated using the same workflow as Section~\ref{sec:data}. For these simulations we also randomly vary the location of the source as well as the velocity model. 

We generate 5000 simulations and from each simulation extract 8 training examples. Each training example contains the previous wavefield, the current wavefield and 11 future wavefields, over different starting time steps.

We use a recursive L2 loss function to train the network. For each training example the network is used to recursively predict 11 time steps ahead, using the output prediction at each time step as the current wavefield input for the next time step. Our L2 loss function is then given by

\begin{equation}
\label{eq:lossrecursive}
L = {1\over N}  \sum_{n+1=1}^{11} \lVert \hat{P}^{n+1} - P^{n+1} \rVert_{2}^{2}~,
\end{equation}

where $N$ is the batch size, $\hat{P}^{n+1}$ is the recursive output prediction of the network at time sample $n+1$ and $P^{n+1}$ is the ground truth wavefield.

In total we extract 20,000 training examples. We use an Adam stochastic gradient descent algorithm with a learning rate of $2\mathrm{x}10^{-4}$ and a batch size of 10 and train over 200,000 steps.

\subsection{\label{sec:none}Results}
Our training loss converges and we assess the performance of our trained network using a validation set of 200 unseen examples. The full wavefield prediction over multiple time steps for 1 randomly selected example in the validation set is shown in Fig.~\ref{fig:wavefields_result}.

For the example shown the trained convolutional network is able to approximate the update equation given by Eq.~\ref{eq:update}. The predicted wavefield expands outward and reflections occur at velocity boundaries. The speed and shape of the wavefront also changes when entering different velocity layers, as expected.

The approximation error of the prediction increases over time (Fig.~\ref{fig:wavefields_result}, bottom). This accumulation of error is likely to occur when recursing through the network multiple times to predict multiple time steps ahead. We find that the recursive loss function given by Eq.~\ref{eq:lossrecursive} reduces but does not remove this cumulative error over time. Similar error accumulation over time was observed by Weiqiang et al. \cite{stanfordwaves}. Unlike Weiqiang et al., we are able to approximate the update equations without the need for multi-scale convolutional networks and using only the current and previous wavefield frames as input.

\section{\label{sec:c}Fast seismic inversion using Wavenet}

\subsection{\label{sec:none}Overview}

Finally, we show a preliminary test for carrying out seismic inversion directly using the WaveNet architecture presented in Section~\ref{sec:a}.

Typically seismic inversion methods such as FWI require an optimisation procedure in order to estimate an Earth model which matches a recorded seismic response. Such methods are not guaranteed to converge and are typically very computationally expensive.

We use the same WaveNet architecture described in Section~\ref{sec:a} to predict a velocity model that satisfies the pressure response recorded at a receiver location in a single inference step, without the need for an optimisation procedure. Such a method could provide a much faster alternative to existing inversion algorithms.

We use the same training data used for our forward prediction network in Section~\ref{sec:a}. Our goal is to train a deep neural network which, given a receiver response as input, can directly estimate a velocity model which satisfies the receiver response.

\subsection{\label{sec:none}Inverse WaveNet architecture}

We use the same prediction workflow as Section~\ref{sec:a} and flip the input and output of the network. We also invert the WaveNet architecture so that the casual correlation between the receiver response and reflectivity series is maintained. In contrast to Fig.~\ref{fig:wavenet}, we only input the single recorded receiver response at zero-offset and the output is a prediction of the normal incidence reflectivity series. We also only use 64 hidden channels for each of the hidden layers in the WaveNet architecture. We recover a prediction of the underlying velocity model by using a standard 1D time-to-depth conversion followed by integration of the reflectivity values.

We use exactly the same training data and training strategy as described in Sections~\ref{sec:data} and \ref{sec:training}. Here our loss function is given by

\begin{equation}
\label{eq:inverse_loss}
L = {1\over N} \lVert \hat{R} - R\rVert_{2}^{2}~,
\end{equation}

where $R$ is the true reflectivity series and $\hat{R}$ is the predicted reflectivity series.

\subsection{\label{sec:none}Results}

Our training loss converges and we test the performance of our trained network using a validation set of 200 unseen examples. The predicted reflectivity series and velocity models for  4 randomly selected examples from this set are shown in Fig.~\ref{fig:inverse_result}.

We see that the WaveNet architecture is able to approximate the inverse function and predict the underlying velocity model for each of these test examples.

Each prediction correctly estimates the number of layers and the sign of each reflectivity spike. The network is able to transform the frequency-limited receiver response into a full-bandwidth sparse reflectivity series.

The time taken for each velocity prediction is similar to the WaveNet prediction times reported in Section~\ref{sec:results}. It is likely our approach is able to make predictions in a fraction of the time needed for seismic inversion algorithms which rely on iterative forward modelling, as we are able to estimate the velocity model in a single inference step.

Due to the integration of the reflectivity series when generating the velocity prediction, small velocity errors propagate in depth in the velocity prediction (for example bottom left, Fig.~\ref{fig:inverse_result}).

\section{\label{sec:c}Conclusions}

We have presented a novel, approximate and fast approach for the simulation of seismic waves. We  designed a  deep  neural  network  which  can simulate  seismic  waves  in  horizontally layered  acoustic  media.  By directly approximating the recorded seismic response at multiple receiver locations we were able to achieve a significant decrease in simulation time compared to traditional FD modelling.

We also proposed novel approaches to full wavefield simulation and for direct seismic inversion using deep learning. In particular our seismic inversion approach potentially offers a much faster method for inversion than traditional inversion techniques.

The speed improvement offered by these networks could lead to real-time seismic simulation applications and could benefit seismic inversion algorithms. Our work suggests that deep learning is a valuable tool for both seismic simulation and inversion. 

Whilst we presented simple proof-of concept approaches here, there is much further work which could be done. We have not yet extended our forward WaveNet network to simulate more arbitrary Earth models. As well as using a WaveNet architecture, attention mechanisms could be tested to focus on the salient parts of the velocity model when predicting each receiver recording \cite{attention}. Anisotropic parameters and other elastic parameters such as shear velocity and density could be given as additional inputs to the network. The network could be extended to 2D and 3D Earth model inputs by increasing its dimensionality. Other alterations such as LSTM or bi-directional RNNs layers could help to improve the  prediction of  multiple reflections. It is not clear whether converting the velocity model to its reflectivity series is appropriate for more complex Earth models. We have yet to test our existing architecture on non-horizontal velocity models and other receiver geometries.

For seismic inversion, our forward WaveNet network is fully differentiable and could therefore be assessed on its ability to produce fast gradient estimates for seismic inversion algorithms such as FWI. Further work could also compare the accuracy of our inverse WaveNet network to traditional seismic inversion. The prediction uncertainty of our networks could be studied using dropout \cite{dropout}. We have yet to test our inverse network on real seismic data.

We finally note that our deep neural networks learned the physics of wave propagation implicitly; no physics knowledge was explicitly coded into the networks.

\begin{acknowledgements}

The authors would like to thank the Computational Infrastructure for Geodynamics (\href{https://www.geodynamics.org/}{www.geodynamics.org}) for releasing the open-source SEISMIC\_CPML FD modelling libraries.

We would also like to thank Tom Le Paine for his fast WaveNet implementation on GitHub from which our code was based on (\href{https://github.com/tomlepaine/fast-wavenet/}{github.com/tomlepaine/fast-wavenet}).

\end{acknowledgements}

\bibliography{main}{}

\end{document}